\documentclass[preprint]{revtex4}

\begin{document}

\title{Resonant Transport in Nb/GaAs/AlGaAs/GaAs Microstructures}

\author{F. Giazotto}
\email{giazotto@sns.it}
\author{P. Pingue}
\author{F. Beltram}
\affiliation{NEST-INFM \& Scuola Normale Superiore, I-56126 Pisa, Italy}

\author{M. Lazzarino}
\author{D. Orani}
\author{S. Rubini}
\author{A. Franciosi}
\affiliation{Laboratorio Nazionale TASC-INFM, Basovizza, I-34012 Trieste, Italy}

\begin{abstract}
Resonant transport in a hybrid semiconductor-superconductor microstructure grown by MBE on GaAs is presented. This structure experimentally realizes the prototype system originally proposed by de Gennes and Saint-James in 1963 in \emph{all}-metal structures. 
A low temperature single  peak superimposed to the characteristic Andreev-dominated subgap conductance represents the mark of such resonant behavior.
Random matrix theory of quantum transport was employed in order to analyze the observed magnetotransport properties and ballistic effects were included by directly solving the Bogoliubov-de Gennes equations.
\end{abstract}
  
\maketitle

In this contribution we report the experimental observation of de Gennes-Saint-James (dGSJ)-type resonances \cite{GJ} in a microstructure consisting of a Nb/GaAs/AlGaAs/GaAs hybrid microjunction.
This microstructure realizes the model system introduced by de Gennes and Saint-James in 1963, consisting of a superconductor-normal metal-insulator-normal metal-type (SNIN) layer sequence.  
Multiple reflections off the superconductive gap (i.e., Andreev reflections \cite{andr}) and off the insulating barrier (i.e., normal reflections) can give rise  to quasi-bound states \cite{GJ,bagwell1} that manifest themselves as conductance resonances.
This result was made possible by the exploitation of semiconductor epitaxial growth to tailor electronic states according to the original dGSJ model system. 

We analyzed the data within the random matrix theory of quantum transport and included ballistic effects by directly solving the Bogoliubov-de Gennes equations in a model potential profile \cite{giazotto}. 
Our description of the system is confirmed by the observed temperature and magnetic-field dependence. 

\section{Samples}

A sketch of the Nb/GaAs/AlGaAs structure is shown in Fig. 1(a). The semiconductor portion consists of a 1 $\mu$m thick $n$-GaAs(001) buffer layer Si-doped nominally at $n= 2\times 10^{18}$ cm$^{-3}$ grown by molecular beam epitaxy (MBE) on a $n$-GaAs(001) substrate, followed by a 4 nm thick Al$_{0.3}$Ga$_{0.7}$As barrier. This was followed by the growth of a 12 nm thick GaAs(001) epilayer Si-doped at $n= 2\times 10^{18}$ cm$^{-3}$ and by 14 nm of GaAs doped by a sequence of six Si $\delta$-doped layers spaced by 2 nm. A 1-$\mu$m-thick amorphous As cap layer was deposited in the MBE growth chamber to protect the surface during transfer in air to an UHV sputter-deposition/surface analysis system. 
After thermal desorption of the As protective cap layer, a 100-nm-thick Nb film was then deposited {\it in situ} by DC-magnetron sputtering.
The thickness of the GaAs epilayer sandwiched between the superconductor and the AlGaAs barrier was selected in order to have an experimentally-accessible single quasi-bound state below the superconductive gap, and the Si $\delta$-doped layers at the Nb/GaAs interface were employed to achieve the required transmissivity. More detail on the contact fabrication procedure is reported elsewhere \cite{nbgaas} where the reference SN junctions are also described. The latter consisted of a Nb/$\delta$-doped-GaAs junction {\it without} the AlGaAs insulating barrier. A qualitative sketch of the energy-band diagram of our SNIN structure is depicted in Fig. 1(b).

100$\times$160\,$\mu$m$^2$ junctions were defined by standard photolithographic techniques and reactive ion etching. 4-wire measurements were performed with two leads on the junction under study and the other two connected to the sample back contact.

\section{Experimental results and discussion}

The measured differential conductances $vs$ bias ($G(V)$) for the resonant structure (SNIN structure, panel (a)) and for the reference junction (SN structure, panel (b)) at $T=0.34$ K are shown in Fig. 2. Comparison of the two characteristics clearly shows the presence of a marked dGSJ subgap conductance peak in the SNIN, resonant device.  The resonance is superimposed to the typical Andreev-dominated subgap conductance. The symmetry of conductance and the zero-bias conductance peak (ZBCP) peculiar to reflectionless tunneling (RT) \cite{nbgaas,kast91,magnee,sanq} further demonstrate the effectiveness of the fabrication protocol.

Quantitative determination of the resonant transport properties in such a system is not trivial as it can be inferred by inspecting Fig. 1(b) and considering the diffusive nature of the normal regions. dGSJ-enhancement, however, is an intrinsically ballistic phenomenon so that its essential features can be captured with relative ease. One study particularly relevant for our system was performed in Ref.\cite{bagwell1}. In the context of ballistic transport a one-dimensional SNIN structure was studied as a function of the N interlayer thickness $d$ and it was demonstrated that resonances can occur in the subgap conductance spectrum for suitable geometric conditions. The insulating barrier  was simulated by a $\delta$-like potential.  Customarily the barrier strength is described by the dimensionless coefficient $Z= {\mathcal V}{_I}/ \hbar v_F$, where $v_F$ is the electron Fermi velocity \cite{btk} and ${\mathcal V}{_I}$ the barrier amplitude, and in the following we shall make use of it to characterize our system. 
The key-results of the analysis are: ({\it a}) the number of resonances increases for larger thickness $d$ of the metallic interlayer and is virtually independent of $Z$; ({\it b}) the energy-width of such resonances decreases for increasing $Z$. 
This model is rather idealized, but is a useful tool to grasp the essential features of our system such as number and position of resonances \cite{bagwell2}. 
In order to apply it we must first determine  parameters such as barrier strength, electron density and mean free path. Also, any quantitative comparison with experiment requires us to estimate the sample series resistance, which influences the experimental energy position of the resonance peak. The above parameters can be obtained from an analysis of the RT-driven ZBCP and from Hall measurements.

We performed Hall measurements at 1.5 K and obtained carrier density $n\simeq 4\times 10^{18}$\,cm$^{-3}$ and mobility $\mu \simeq 1.5\times 10^{3}$\,cm$^{2}$/V s. These data allow us to estimate the thermal coherence length $L_T =\sqrt{\hbar D/2\pi k_{B}T}=\,0.13\,\mu$m$/\sqrt{T}$, where $D=1.34\times 10^2$\,cm$^2$/s is the diffusion constant, and the electron mean free path ${\ell}_{m}\,\approx 48$\,nm.  $ {\ell}_{m}$ compares favorably with the geometrical constraints of the structure ($d=26$ nm $< {\ell}_{m}$) and further supports our ballistic analysis.

The ZBCP can be described following the work by Beenakker and co-workers  \cite{marmor}. Their model system consists of a normal disordered region  of length $L$ and width $W$ separated by a finite-transparency  barrier from a superconductor. Under appropriate conditions a ZBCP is predicted to occur, whose width $V_c$ (i.e., the voltage at which the ZBCP is suppressed) was estimated to be of the order of the Thouless voltage, $V_c =\hbar D/e L^2$. Similarly, upon application of a magnetic field, the ZBCP is suppressed for fields of the order of $B_c=h/eLW$. In real systems, at finite temperatures, $L$ and $W$ are to be replaced by $L_T$, if it is smaller \cite{marmor}. 
At 0.34 K $L_T \approx 0.22$\,$\mu$m and in our junctions we calculate $V_c \simeq 200$\,$\mu$V. Comparison with the experimental value $V_c^{exp} \simeq 600$\,$\mu$V in Fig. 2(c) (see solid line), allowed us to estimate the series resistance contribution to the measured conductivity. This rather large effect stems mainly from the AlGaAs barrier and the sample back-contact resistance.

We calculated the conductance for the nominal N-thickness value ($d = 26$ nm) and $Z = 1.4$, as appropriate for the AlGaAs barrier \cite{giazotto,quantiz}. We emphasize, however, that the essential features such as the number and energy position of the dGSJ resonances are virtually independent of Z and are controlled instead by the value of $d$. Our calculations yield a single conductance peak at energies corresponding to about $0.8\,\Delta$, where $\Delta$ is the superconductor energy gap. By including the above-determined series resistance contribution, the resonance peak is positioned at about $3.5\,$mV (see Fig. 2(c), dash-dotted line, $T=0.34$ K). 
The corresponding experimental value is about 3 mV (solid line in Fig. 2(c)), but the observed energy difference is well within the uncertainty resulting from the determination of the series resistance.
The results of our model calculations strongly support our interpretation of the experimental structure in terms of dGSJ resonant transport.

Ordinary resonant tunneling in the normal double-barrier potential
schematically shown in Fig. 1(b) cannot explain the observed subgap structure. This is indicated
by the symmetry in the experimental data for positive and negative bias and is further proven
by the temperature and magnetic field dependence of the differential conductance. 
Figure 3 shows a set of $G(V)$s measured in the 0.34-1.55 K range for the ZBCP, (a), and for the resonance peak, (b). Both effects show a strong dependence on temperature and at $T=1.55$ K are totally suppressed. This temperature value is within the range where RT suppression is expected \cite{kast91,magnee,sanq,nbgaas}.
At higher temperatures, the conductance in the resonance region resembles that of the reference SN junction of Fig. 2(b).
Notably, the ZBCP and the resonance peak disappear at the same temperature, hence indicating the coherent nature of the observed effect.

Further confirmation of the nature of the resonance peak can be gained observing its dependence from the magnetic field. Figure 4 shows $G(V)$ at $T=0.35$\,K for several values of the magnetic field applied in the plane of the junction for the ZBCP, (a), and for the resonance peak, (b). 
The measurements confirm the known sensitivity of ZBCP to the magnetic field \cite{marmor}, and clearly indicate that the dGSJ resonance is easily suppressed for critical fields of the order of 100 mT. 
In the perpendicular field configuration at $T=0.36$\,K (see Fig. 4(c),(d)) 
 the resonance and the ZBCP displayed a similar dependence. Such behavior is fully consistent with dGSJ-related origin but is not compatible with a normal resonant tunneling description of the data \cite{rtsemicond}.  

\section{Conclusions}
In summary, we have experimentally observed dGSJ resonant states in Nb/GaAs/AlGaAs hybrid heterostructures. 
Transport was studied as a function of temperature and magnetic field and was successfully described within the ballistic model of Riedel and Bagwell\cite{bagwell1}.
The present results suggest that the Nb/GaAs/AlGaAs system is a good candidate for the implementation of complex mesoscopic structures that can take advantage of the mature AlGaAs nanofabrication technology. Such structures may represent ideal prototype systems for the study of coherent transport and the implementation of novel hybrid devices.

\section*{Acknowledgments}
This work was sponsored by INFM under the PAIS project EISS.

%------ Figures beginning -------------

\begin{figure}
%\begin{center}
%\includegraphics[width=4.5cm,angle=-90]{Fig1.eps}
%\figurebox{10pc}{15pc}{kkk.ps} % to have a box alone
%\epsfxsize=12pc % will enlarge or reduce the postscript figures based on the xsize
%\epsfbox{Fig1bis.eps} % postscript image file name
%
\caption{(a) Schematic structure of the Nb/GaAs/AlGaAs system  analyzed in this work. (b) Sketch of the  energy-band diagram of the structure. The shaded area represents de Gennes-Saint-James quasi-bound state confined between the superconductor and the Al$_{0.3}$Ga$_{0.7}$As barrier.}  
\end{figure}

\begin{figure}
%\begin{center}
%\includegraphics[width=6.7cm,angle=-90]{Fig2.eps}
%\figurebox{10pc}{15pc}{} % to have a box alone
%\epsfxsize=15pc % will enlarge or reduce the postscript figures based on the xsize
%\epsfbox{Fig2.eps} % postscript image file name
%\end{center}
\caption{Differential conductance $vs$ voltage at $T=0.34$ K for the SNIN junction, (a), and for the reference SN junction, (b). Both curves show  reflectionless tunneling enhancement around zero bias, while only in (a) a finite-bias, subgap de Gennes-Saint-James peak is present. (c) Comparison at $T=0.34$ K between the experimental differential conductance $G(V)$ (solid line) and the numerical simulation obtained following the model of Riedel and Bagwell (dash-dotted line, see text).}
\end{figure}

\begin{figure}
%\begin{center}
%\includegraphics[width=6cm,angle=-90]{Fig3.eps}
%\figurebox{10pc}{15pc}{} % to have a box alone
%\epsfxsize=15pc % will enlarge or reduce the postscript figures based on the xsize
%\epsfbox{Fig3.eps} % postscript image file name
%\end{center}
\caption{ Differential conductance $vs$ voltage at several temperatures in the 0.34 to 1.55\,K range (a) for the zero bias conductance peak, and (b) for the de Gennes-Saint-James resonance peak. 
Data were taken at: (a) $T=0.34,0.45,0.55,0.65,0.75,0.85,0.95,1.05,1.26,1.45,1.50,1.55$ K; (b) $T=0.34,0.55,0.65,0.85,1.05,1.26,1.55$ K.
}
\end{figure}

\begin{figure}
%\begin{center}
%\includegraphics[width=7.5cm,angle=-90]{Fig6.eps}
%\figurebox{10pc}{15pc}{} % to have a box alone
%\epsfxsize=15pc % will enlarge or reduce the postscript figures based on the xsize
%\epsfbox{Fig3.eps} % postscript image file name
%\end{center}
\caption{ Differential conductance $vs$ voltage for several values of the magnetic field applied in-plane ((a),(b), $T=0.35$\,K) and perpendicularly ((c),(d), $T=0.36$\,K) to the junction. Dependence ((a),(c)) of the  zero bias conductance, and ((b),(d)) of the de Gennes-Saint-James resonance peak. In (b) data were taken at $H_{//}=0,10,20,25,30,35,40,45,50,60,70,80,100,150$ mT.
}
\end{figure}


\begin{references}
\bibitem{quantiz} Taking into account the Al$_{0.3}$Ga$_{0.7}$As/GaAs discontinuity, Fermi energy and barrier thickness it is straightforward to determine $Z\simeq 1.4$.
\end{references}
\end{document}